\documentclass[manuscript]{acmart}

\AtBeginDocument{%
  \providecommand\BibTeX{{%
    \normalfont B\kern-0.5em{\scshape i\kern-0.25em b}\kern-0.8em\TeX}}}

\setcopyright{acmlicensed}
\copyrightyear{2024}
\acmYear{2024}
\acmDOI{}
\acmConference[XAIxArts 2024]{Explainable AI for the Arts Workshop 2024}{June 23, 2024}{Chicago, IL, United States}

\begin{document}

\title{A Mapping Strategy for Interacting with Latent Audio Synthesis Using Artistic Materials}


\author{Shuoyang Zheng}
\email{shuoyang.zheng@qmul.ac.uk}
\orcid{0000-0002-5483-6028}
\affiliation{%
  \institution{Centre for Digital Music, Queen Mary University of London}
  \country{UK}
}

\author{Anna Xambó Sedó}
\email{a.xambosedo@qmul.ac.uk}
\orcid{0000-0003-2333-6941}
\affiliation{%
  \institution{Centre for Digital Music, Queen Mary University of London}
  \country{UK}
}
\author{Nick Bryan-Kinns}
\email{n.bryankinns@arts.ac.uk}
\orcid{0000-0002-1382-2914}
\affiliation{%
  \institution{Creative Computing Institute, University of the Arts London}
  \country{UK}
}

\makeatletter
\let\@authorsaddresses\@empty
\makeatother

\renewcommand{\shortauthors}{Zheng et al.}

\begin{abstract}
This paper presents a mapping strategy for interacting with the latent spaces of generative AI models. Our approach involves using unsupervised feature learning to encode a human control space and mapping it to an audio synthesis model's latent space. To demonstrate how this mapping strategy can turn high-dimensional sensor data into control mechanisms of a deep generative model, we present a proof-of-concept system that uses visual sketches to control an audio synthesis model. We draw on emerging discourses in XAIxArts to discuss how this approach can contribute to XAI in artistic and creative contexts, we also discuss its current limitations and propose future research directions.
\end{abstract}


\maketitle

\section{Introduction}
Research in the ﬁeld of eXplainable AI for the Arts (XAIxArts) examines how AI models could be made more understandable for creative activities \cite{bryan-kinns_reflections_2024}. One approach to increasing the explainability of an AI model is to provide real-time interaction and feedback \cite{bryan-kinns_exploring_2021}. This approach allows the user to learn how the model works and find ways to interact with it by exploring how their actions affect the model's output \cite{murray-browne_latent_2021}. In practice, it involves mapping a human control space to the parameter space of the AI model and providing real-time audio feedback. For example, mapping a user's spatial coordination in an immersive environment to the latent space of a deep audio synthesis model so that the user learns the effect of specific latent space axes by `walking through' the environment \cite{scurto_soundwalking_2023}. 

The design of this human control space is an open question in the field of XAI \cite{bryan-kinns_explainable_2023}, especially in a creative and artistic context. While typical approaches have explored parametric control mechanisms such as sliders and 2D pads \cite{vigliensoni_steering_2023, bryan-kinns_exploring_2021}, it is important to consider other interaction mediums that may be more expressive in creative and artistic contexts \cite{clemens_explaining_2023}. These interaction mediums can be the artistic materials from the artists' performance space \cite{bryan-kinns_reflections_2024}, for example, the movements of dancers \cite{murray-browne_latent_2021}, sketches \cite{lobbers_sketchsynth_2023}, or electromyography data of performers \cite{silva_interactive_2020}. However, materials in an artistic performance space can be high-dimensional sensor data, which are difficult to map to the AI model and use as a medium for interaction \cite{roma_adaptive_2019}. To tackle this, we propose a mapping strategy that encodes a human performance space and maps it to the latent space of an AI model to help users interact with the model's generation process. We use a sketch-to-sound controller we developed in a previous work \cite{zheng_building_2024} to demonstrate how this mapping strategy can turn high-dimensional sensor data into control mechanisms of an AI model. 

\section{Mapping Strategy}

\begin{figure}[htbp]
	\centering
		\includegraphics[width=0.7\columnwidth]{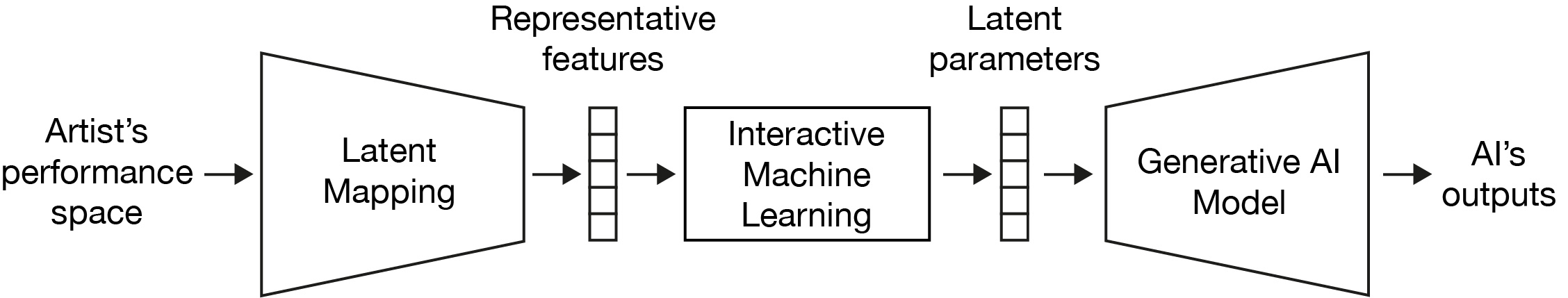}
	\caption{Proposed mapping strategy}
	\label{fig:mapping}
\end{figure}

Materials in an artistic performance space can be high-dimensional sensor data such as sketches \cite{lobbers_sketchsynth_2023} or movements of dancers \cite{murray-browne_latent_2021}. In order to map artistic materials to the latent space of an AI model, a two-step approach is needed: (i) feature extraction techniques are required to encode these data into representative features \cite{roma_adaptive_2019}, (ii) these representative features need to be meaningfully mapped to the latent space of the AI model. Therefore, as shown in Figure~\ref{fig:mapping}, we propose using latent mapping \cite{murray-browne_latent_2021} and Interactive Machine Learning (IML) \cite{fiebrink_meta-instrument_2009} to achieve these two steps.

\textbf{Step 1} uses latent mapping \cite{murray-browne_latent_2021}, an unsupervised feature learning approach, to learn representative features from a large corpus of unlabelled data. A trained latent mapping model encodes high-dimensional inputs into lower-dimensional latent
representations, holding representative features of the artist's current movements, sketches, or other artistic practices, which can be mapped to sound controls. Commonly used unsupervised feature learning methods include Variational AutoEncoders \cite{murray-browne_latent_2021} and unsupervised dimensionality reduction \cite{roma_adaptive_2019}. In practice, training a latent mapping model requires the artist to create a training dataset for their artistic materials. In other words, they need to record a full set of possible outputs that they might produce in the performance space \cite{murray-browne_latent_2021}. For example, training a latent mapping model for a dancer may require recording body skeletal data from the dancer of all their possible dance movements, including intermediate transitions between two movements. 

\textbf{Step 2} uses Interactive Machine Learning (IML) to connect the extracted features and the AI model's parameters. IML allows users to construct mappings between two spaces using a few personalised training examples \cite{fiebrink_meta-instrument_2009}. In this way, it creates a mapping between the human performance space and the AI model's output, allowing artists to interact with the AI model using data from the performance space. In practice, the artist manually defines a few paired data points in the performance space and the AI model's parameter space as training examples. Tools such as {\itshape Wekinator} \cite{fiebrink_meta-instrument_2009} can be used to facilitate this process.

\section{Demonstration}

To demonstrate how this mapping strategy works in practice, we build on our previous work on a sketch-to-sound controller that encodes visual sketches into sound controls. In this work, we use it as the latent mapping model to control a latent audio synthesis model. Technical details for the sketch-to-sound controller can be found in \cite{zheng_building_2024}. We use RAVE \cite{caillon_rave_2021}, a variational autoencoder that synthesises audio from a latent space in real-time. We use a RAVE model pre-trained on organ music recordings \cite{intelligent_instruments_lab_rave-models_2023} that runs in a Max4Live\footnote{\url{https://maxforlive.com/library/}} device. The pre-trained model has 16 latent space dimensions, and the sketch-to-sound model encodes sketches into a 32-dimensional latent space. Therefore, the IML model takes a 32-dimensional input and maps it to a 16-dimensional output. The communications between the sketch-to-sound model, the IML model, and the RAVE model are achieved through open sound control (OSC)\footnote{\url{https://ccrma.stanford.edu/groups/osc/index.html}}. 

\begin{figure}[tbp]
	\centering
		\includegraphics[width=\columnwidth]{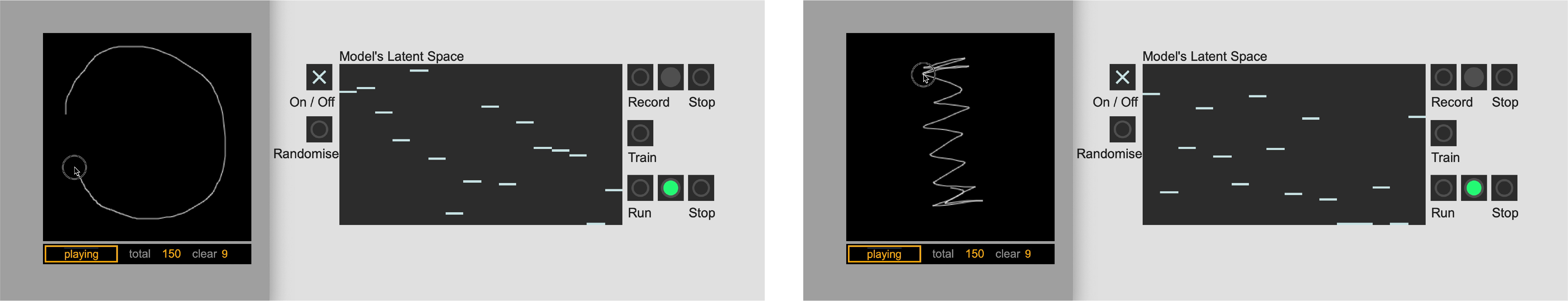}
	\caption{Two screenshots of user interfaces showing the sketch and the audio synthesis model's latent spaces. A full demonstration video can be viewed at \url{https://tinyurl.com/xaixarts24-mapping}.}
	\label{fig:interface}
\end{figure}

Figure~\ref{fig:interface} shows two screenshots of the user interface. In each screenshot, the left panel allows the user to interact with the RAVE audio synthesis model by sketching on the canvas, while the right panel allows users to train the IML model. When recording training examples for the IML model, users can start by either manually adjusting the latent parameters or randomising the RAVE model's latent parameters using the ``randomise'' button, then clicking the ``record'' button and drawing sketches that depict the current sound produced by the RAVE model. After training the IML model, users can click the ``run'' button and interact with the RAVE model by drawing sketches.

\section{Limitations and Future Work}
This section draws on emerging discourses in XAIxArts \cite{bryan-kinns_explainable_2023} to discuss the proposed mapping strategy in terms of the temporal and cross-modal aspects. It also discusses the limitations and contributes with future research directions.

First, \textbf{the temporal aspect}. Real-time interaction and feedback in an XAI model can help artists discover and reuse unexpected phenomena for their artistic expression \cite{bryan-kinns_reflections_2024}. For example, in our demonstration video, movements such as quick transitions from short lines to full circles trigger interesting sounds. Users can reuse these sounds by memorising the sketch trajectories that trigger them, but is there a way to explain these trajectories to the users in the first place? Typical XAI approaches focus on explaining a static state of the model parameters \cite{dhanorkar_who_2021}, but explaining the trajectories of sketches focuses on explaining the transition, or the interpolation from one state to another, which involves taking account of the temporal dimension. This aspect of explanation contributes to the embodied knowledge about the AI model because it tells artists how to act with a system, which is an important component in artistic and creative contexts \cite{murray-browne_emergent_2021}.

Second, \textbf{the cross-modal aspect}. Investigating interactions that involve cross-modalities, such as shape-sound associations, is useful to expand the explanation mediums in creative and artistic contexts \cite{clemens_explaining_2023}. In latent audio models, for example, future research can explore how human-interpretable features in non-audio domains can be applied to audio latent space. Typical XAI approaches aim to establish meaningful features in the audio domain, and a number of these features already rely on cross-modal metaphors to express the sonic characteristics, such as ``brightness'', ``noisiness'', and so on \cite{lobbers_sketchsynth_2023}. Our mapping strategy can be used to study other cross-modal mappings that might be useful in XAI. For example, future work on our sketch-to-sound system can explore programming and animating sketches using interpretable features such as rotation, zoom, or other affine transformations, and then use our mapping strategy to connect these features with the audio latent space to discover meaningful shape-sound associations that can be used to explain the audio synthesis model.

The motivation for the mapping strategy presented in this paper was to encourage explorations on explanation mediums that are salient in the creative and artistic contexts \cite{clemens_explaining_2023}. However, we only use sketch-to-sound to demonstrate how the mapping strategy works. It would be necessary to explore other forms of creative practices and go beyond sketch and sound to generalise the discussion into wider creative and artistic contexts.

\section{Conclusion}

Real-time interaction and feedback increase the explainability of an AI model by supporting users in exploring how their actions affect the audio generation. Following this direction, this paper proposed a mapping strategy for interacting with latent audio synthesis models using data from an artistic performance space. We have described how this mapping strategy works in practice and demonstrated it with a system that controls an AI audio synthesis model using visual sketches. We discussed that future work using this mapping strategy should take account of the temporal and cross-modal XAI aspects, and it is important to experiment with how this mapping strategy can be used in other forms of creative practices in future work.

\begin{acks}
Shuoyang Zheng is a research student at the UKRI Centre for Doctoral Training in Artificial Intelligence and Music, supported by UK Research and Innovation [grant number EP/S022694/1].
\end{acks}

\bibliographystyle{ACM-Reference-Format}
\bibliography{sample-base}


\begin{thebibliography}{16}


\ifx \showCODEN    \undefined \def \showCODEN     #1{\unskip}     \fi
\ifx \showDOI      \undefined \def \showDOI       #1{#1}\fi
\ifx \showISBNx    \undefined \def \showISBNx     #1{\unskip}     \fi
\ifx \showISBNxiii \undefined \def \showISBNxiii  #1{\unskip}     \fi
\ifx \showISSN     \undefined \def \showISSN      #1{\unskip}     \fi
\ifx \showLCCN     \undefined \def \showLCCN      #1{\unskip}     \fi
\ifx \shownote     \undefined \def \shownote      #1{#1}          \fi
\ifx \showarticletitle \undefined \def \showarticletitle #1{#1}   \fi
\ifx \showURL      \undefined \def \showURL       {\relax}        \fi
\providecommand\bibfield[2]{#2}
\providecommand\bibinfo[2]{#2}
\providecommand\natexlab[1]{#1}
\providecommand\showeprint[2][]{arXiv:#2}

\bibitem[Bryan-Kinns(2024)]%
        {bryan-kinns_reflections_2024}
\bibfield{author}{\bibinfo{person}{Nick Bryan-Kinns}.} \bibinfo{year}{2024}\natexlab{}.
\newblock \showarticletitle{Reflections on {Explainable} {AI} for the {Arts} ({XAIxArts})}.
\newblock \bibinfo{journal}{\emph{Interactions}} \bibinfo{volume}{31}, \bibinfo{number}{1} (\bibinfo{date}{Jan.} \bibinfo{year}{2024}), \bibinfo{pages}{43--47}.
\newblock
\showISSN{1072-5520}
\urldef\tempurl%
\url{https://doi.org/10.1145/3636457}
\showDOI{\tempurl}


\bibitem[Bryan-Kinns et~al\mbox{.}(2021)]%
        {bryan-kinns_exploring_2021}
\bibfield{author}{\bibinfo{person}{Nick Bryan-Kinns}, \bibinfo{person}{Berker Banar}, \bibinfo{person}{Corey Ford}, \bibinfo{person}{Courtney~N Reed}, \bibinfo{person}{Yixiao Zhang}, \bibinfo{person}{Simon Colton}, {and} \bibinfo{person}{Jack Armitage}.} \bibinfo{year}{2021}\natexlab{}.
\newblock \showarticletitle{Exploring {XAI} for the {Arts}: {Explaining} {latent} {space} in {generative} {music}}. In \bibinfo{booktitle}{\emph{1st {Workshop} on {eXplainable} {AI} approaches for debugging and diagnosis}}.
\newblock


\bibitem[Bryan-Kinns et~al\mbox{.}(2023)]%
        {bryan-kinns_explainable_2023}
\bibfield{author}{\bibinfo{person}{Nick Bryan-Kinns}, \bibinfo{person}{Corey Ford}, \bibinfo{person}{Alan Chamberlain}, \bibinfo{person}{Steven~David Benford}, \bibinfo{person}{Helen Kennedy}, \bibinfo{person}{Zijin Li}, \bibinfo{person}{Wu Qiong}, \bibinfo{person}{Gus~G. Xia}, {and} \bibinfo{person}{Jeba Rezwana}.} \bibinfo{year}{2023}\natexlab{}.
\newblock \showarticletitle{Explainable {AI} for the {Arts}: {XAIxArts}}. In \bibinfo{booktitle}{\emph{Creativity and {Cognition}}}. \bibinfo{publisher}{ACM}, \bibinfo{address}{Virtual Event USA}, \bibinfo{pages}{1--7}.
\newblock
\showISBNx{9798400701801}
\urldef\tempurl%
\url{https://doi.org/10.1145/3591196.3593517}
\showDOI{\tempurl}


\bibitem[Caillon and Esling(2021)]%
        {caillon_rave_2021}
\bibfield{author}{\bibinfo{person}{Antoine Caillon} {and} \bibinfo{person}{Philippe Esling}.} \bibinfo{year}{2021}\natexlab{}.
\newblock \showarticletitle{{RAVE}: {A} variational autoencoder for fast and high-quality neural audio synthesis}.
\newblock  (\bibinfo{year}{2021}).
\newblock
\urldef\tempurl%
\url{https://arxiv.org/abs/2111.05011}
\showURL{%
\tempurl}
\newblock
\shownote{arXiv: 2111.05011}.


\bibitem[Clemens(2023)]%
        {clemens_explaining_2023}
\bibfield{author}{\bibinfo{person}{Michael Clemens}.} \bibinfo{year}{2023}\natexlab{}.
\newblock \showarticletitle{Explaining the {arts}: {Toward} a {framework} for {matching} {creative} {tasks} with {appropriate} {explanation} {mediums}}. In \bibinfo{booktitle}{\emph{Proceedings of {The} first international workshop on {eXplainable} {AI} for the {Arts} ({XAIxArts})}}.
\newblock


\bibitem[Dhanorkar et~al\mbox{.}(2021)]%
        {dhanorkar_who_2021}
\bibfield{author}{\bibinfo{person}{Shipi Dhanorkar}, \bibinfo{person}{Christine~T. Wolf}, \bibinfo{person}{Kun Qian}, \bibinfo{person}{Anbang Xu}, \bibinfo{person}{Lucian Popa}, {and} \bibinfo{person}{Yunyao Li}.} \bibinfo{year}{2021}\natexlab{}.
\newblock \showarticletitle{Who needs to know what, when?: {Broadening} the {Explainable} {AI} ({XAI}) {Design} {Space} by {Looking} at {Explanations} {Across} the {AI} {Lifecycle}}. In \bibinfo{booktitle}{\emph{Proceedings of the 2021 {ACM} {Designing} {Interactive} {Systems} {Conference}}} \emph{(\bibinfo{series}{{DIS} '21})}. \bibinfo{publisher}{Association for Computing Machinery}, \bibinfo{address}{New York, NY, USA}, \bibinfo{pages}{1591--1602}.
\newblock
\showISBNx{978-1-4503-8476-6}
\urldef\tempurl%
\url{https://doi.org/10.1145/3461778.3462131}
\showDOI{\tempurl}


\bibitem[Fiebrink et~al\mbox{.}(2009)]%
        {fiebrink_meta-instrument_2009}
\bibfield{author}{\bibinfo{person}{Rebecca Fiebrink}, \bibinfo{person}{Dan Trueman}, {and} \bibinfo{person}{Perry~R. Cook}.} \bibinfo{year}{2009}\natexlab{}.
\newblock \showarticletitle{A {meta}-{instrument} {for} {interactive}, {on}-{the}-{fly} {machine} {learning}}. In \bibinfo{booktitle}{\emph{Proceedings of the {International} {Conference} on {New} {Interfaces} for {Musical} {Expression}}}. \bibinfo{publisher}{Zenodo}, \bibinfo{address}{Pittsburgh, PA, United States}, \bibinfo{pages}{280--285}.
\newblock
\urldef\tempurl%
\url{https://zenodo.org/record/1177513}
\showURL{%
\tempurl}


\bibitem[{Intelligent Instruments Lab}(2023)]%
        {intelligent_instruments_lab_rave-models_2023}
\bibfield{author}{\bibinfo{person}{{Intelligent Instruments Lab}}.} \bibinfo{year}{2023}\natexlab{}.
\newblock \bibinfo{title}{rave-models ({Revision} ad15daf)}.
\newblock
\newblock
\urldef\tempurl%
\url{https://doi.org/10.57967/hf/1235}
\showDOI{\tempurl}


\bibitem[Lobbers and Fazekas(2023)]%
        {lobbers_sketchsynth_2023}
\bibfield{author}{\bibinfo{person}{Sebastian Lobbers} {and} \bibinfo{person}{George Fazekas}.} \bibinfo{year}{2023}\natexlab{}.
\newblock \showarticletitle{{SketchSynth}: A browser-based sketching interface for sound control}.
\newblock \bibinfo{journal}{\emph{Proceedings of the International Conference on New Interfaces for Musical Expression}} (\bibinfo{date}{May} \bibinfo{year}{2023}), \bibinfo{pages}{637--641}.
\newblock
\showISSN{2220-4806}
\urldef\tempurl%
\url{http://nime.org/proceedings/2023/nime2023_95.pdf}
\showURL{%
\tempurl}


\bibitem[Murray-Browne and Tigas(2021a)]%
        {murray-browne_emergent_2021}
\bibfield{author}{\bibinfo{person}{Tim Murray-Browne} {and} \bibinfo{person}{Panagiotis Tigas}.} \bibinfo{year}{2021}\natexlab{a}.
\newblock \showarticletitle{Emergent {interfaces}: {Vague}, {complex}, {bespoke} and {embodied} {interaction} between {humans} and {computers}}.
\newblock \bibinfo{journal}{\emph{Applied Sciences}} \bibinfo{volume}{11}, \bibinfo{number}{18} (\bibinfo{year}{2021}).
\newblock
\showISSN{2076-3417}
\urldef\tempurl%
\url{https://doi.org/10.3390/app11188531}
\showDOI{\tempurl}


\bibitem[Murray-Browne and Tigas(2021b)]%
        {murray-browne_latent_2021}
\bibfield{author}{\bibinfo{person}{Tim Murray-Browne} {and} \bibinfo{person}{Panagiotis Tigas}.} \bibinfo{year}{2021}\natexlab{b}.
\newblock \showarticletitle{Latent {mappings}: {Generating} {open}-{ended} {expressive} {mappings} {using} {variational} {autoencoders}}. In \bibinfo{booktitle}{\emph{International {Conference} on {New} {Interfaces} for {Musical} {Expression}}}.
\newblock
\urldef\tempurl%
\url{https://doi.org/10.21428/92fbeb44.9d4bcd4b}
\showDOI{\tempurl}


\bibitem[Roma et~al\mbox{.}(2019)]%
        {roma_adaptive_2019}
\bibfield{author}{\bibinfo{person}{Gerard Roma}, \bibinfo{person}{Owen Green}, {and} \bibinfo{person}{Pierre~Alexandre Tremblay}.} \bibinfo{year}{2019}\natexlab{}.
\newblock \showarticletitle{Adaptive {mapping} of {sound} {collections} for {data}-driven {musical} {interfaces}}. In \bibinfo{booktitle}{\emph{Proceedings of the {International} {Conference} on {New} {Interfaces} for {Musical} {Expression}}}, \bibfield{editor}{\bibinfo{person}{Marcelo Queiroz} {and} \bibinfo{person}{Anna~Xambó Sedó}} (Eds.). \bibinfo{publisher}{UFRGS}, \bibinfo{address}{Porto Alegre, Brazil}, \bibinfo{pages}{313--318}.
\newblock
\urldef\tempurl%
\url{https://doi.org/10.5281/zenodo.3672976}
\showDOI{\tempurl}
\newblock
\shownote{ISSN: 2220-4806}.


\bibitem[Scurto and Postel(2023)]%
        {scurto_soundwalking_2023}
\bibfield{author}{\bibinfo{person}{Hugo Scurto} {and} \bibinfo{person}{Ludmila Postel}.} \bibinfo{year}{2023}\natexlab{}.
\newblock \showarticletitle{Soundwalking {deep} {latent} {spaces}}. In \bibinfo{booktitle}{\emph{Proceedings of the 23rd {International} {Conference} on {New} {Interfaces} for {Musical} {Expression} ({NIME}'23)}}. \bibinfo{address}{Mexico, Mexico}.
\newblock
\urldef\tempurl%
\url{https://hal.science/hal-04108997}
\showURL{%
\tempurl}


\bibitem[Silva et~al\mbox{.}(2020)]%
        {silva_interactive_2020}
\bibfield{author}{\bibinfo{person}{Hugo Silva}, \bibinfo{person}{Michael Zbyszynski}, \bibinfo{person}{Atau Tanaka}, {and} \bibinfo{person}{Federico Visi}.} \bibinfo{year}{2020}\natexlab{}.
\newblock \bibinfo{title}{Interactive {machine} {learning}: {Strategies} for live performance using {electromyography}}.  (\bibinfo{year}{2020}).
\newblock
\urldef\tempurl%
\url{https://research.gold.ac.uk/id/eprint/28215/}
\showURL{%
\tempurl}


\bibitem[Vigliensoni and Fiebrink(2023)]%
        {vigliensoni_steering_2023}
\bibfield{author}{\bibinfo{person}{Gabriel Vigliensoni} {and} \bibinfo{person}{Rebecca Fiebrink}.} \bibinfo{year}{2023}\natexlab{}.
\newblock \showarticletitle{Steering latent audio models through interactive machine learning}. In \bibinfo{booktitle}{\emph{In {Proceedings} of the 14th {International} {Conference} on {Computational} {Creativity}}}. \bibinfo{address}{Ontario, Canada}.
\newblock
\urldef\tempurl%
\url{https://computationalcreativity.net/iccc23/papers/ICCC-2023_paper_100.pdf}
\showURL{%
\tempurl}


\bibitem[Zheng et~al\mbox{.}(pted)]%
        {zheng_building_2024}
\bibfield{author}{\bibinfo{person}{Shuoyang Zheng}, \bibinfo{person}{Bleiz~M. Del~Sette}, \bibinfo{person}{Charalampos Saitis}, \bibinfo{person}{Anna Xambó}, {and} \bibinfo{person}{Nick Bryan-Kinns}.} \bibinfo{year}{Accepted}\natexlab{}.
\newblock \showarticletitle{Building {sketch}-to-{sound} {mapping} with {unsupervised} {feature} {extraction} and {interactive} {machine} {learning}}.
\newblock \bibinfo{journal}{\emph{Proceedings of the International Conference on New Interfaces for Musical Expression}} (\bibinfo{year}{Accepted}), \bibinfo{pages}{Utrecht, Netherlands}.
\newblock


\end{thebibliography}

\end{document}